\documentclass[
 reprint,
 superscriptaddress,
 amsmath,
 amssymb,
 aps,
 prl,
]{revtex4-2}

\usepackage{graphicx}% Include figure files
\usepackage{dcolumn}% Align table columns on decimal point
\usepackage{bm}% bold math
\usepackage{times}
\usepackage{xcolor}
\usepackage{hyperref}
\hypersetup{hypertex=true, 
	colorlinks=true, 
	linkcolor=blue, 
	anchorcolor=blue, 
	citecolor=blue}
\begin{document}

%\preprint{APS/123-QED}

\title{Multi-Dressed-State Engineered Rydberg Electrometry}% Force line breaks with \\

\author{Yuhan Yan}
\affiliation{%
	State Key Laboratory of Precision Spectroscopy, Institute of Quantum Science and Precision Measurement, School of Physics, East China Normal University, Shanghai, 200062, China
}%
\affiliation{%
	Wangzhijiang Innovation Center for Laser,  Aerospace Laser Technology and System Department, Shanghai Institute of Optics and Fine Mechanics, Chinese Academy of Sciences, Shanghai, 201800, China
}%
\author{Bowen Yang}
\affiliation{%
	Wangzhijiang Innovation Center for Laser,  Aerospace Laser Technology and System Department, Shanghai Institute of Optics and Fine Mechanics, Chinese Academy of Sciences, Shanghai, 201800, China
}%
\affiliation{%
	Center of Materials Science and Optoelectronics Engineering, University of Chinese Academy of Sciences, Beijing, 100049, China
}%
\author{Xuejie Li}
\affiliation{%
	State Key Laboratory of Precision Spectroscopy, Institute of Quantum Science and Precision Measurement, School of Physics, East China Normal University, Shanghai, 200062, China
}%
\affiliation{%
	Wangzhijiang Innovation Center for Laser,  Aerospace Laser Technology and System Department, Shanghai Institute of Optics and Fine Mechanics, Chinese Academy of Sciences, Shanghai, 201800, China
}%
\author{Haojie Zhao}
\affiliation{%
	Wangzhijiang Innovation Center for Laser,  Aerospace Laser Technology and System Department, Shanghai Institute of Optics and Fine Mechanics, Chinese Academy of Sciences, Shanghai, 201800, China
}%
\affiliation{%
	Center of Materials Science and Optoelectronics Engineering, University of Chinese Academy of Sciences, Beijing, 100049, China
}%
\author{Binghong Yu}
\affiliation{%
	Wangzhijiang Innovation Center for Laser,  Aerospace Laser Technology and System Department, Shanghai Institute of Optics and Fine Mechanics, Chinese Academy of Sciences, Shanghai, 201800, China
}%
\affiliation{%
	Center of Materials Science and Optoelectronics Engineering, University of Chinese Academy of Sciences, Beijing, 100049, China
}%
\author{Jianliao Deng}
\altaffiliation{jldeng@siom.ac.cn}
\affiliation{%
	Wangzhijiang Innovation Center for Laser,  Aerospace Laser Technology and System Department, Shanghai Institute of Optics and Fine Mechanics, Chinese Academy of Sciences, Shanghai, 201800, China
}%
\affiliation{%
	Center of Materials Science and Optoelectronics Engineering, University of Chinese Academy of Sciences, Beijing, 100049, China
}%
\author{L. Q. Chen}
\altaffiliation{lqchen@phy.ecnu.edu.cn}
\affiliation{%
	State Key Laboratory of Precision Spectroscopy, Institute of Quantum Science and Precision Measurement, School of Physics, East China Normal University, Shanghai, 200062, China
}%
\affiliation{%
	Hefei National Laboratory, Hefei, 230088, China
}%
\author{Huadong Cheng}
\altaffiliation{chenghd@siom.ac.cn}
\affiliation{%
	Wangzhijiang Innovation Center for Laser,  Aerospace Laser Technology and System Department, Shanghai Institute of Optics and Fine Mechanics, Chinese Academy of Sciences, Shanghai, 201800, China
}%
\affiliation{%
	Center of Materials Science and Optoelectronics Engineering, University of Chinese Academy of Sciences, Beijing, 100049, China
}%


\raggedbottom
\begin{abstract}
Rydberg atoms, with their giant transition electric dipole moments and abundant
energy-level transitions, offer exceptional potential for microwave (MW)
electric field sensing, combining high sensitivity and broad frequency
coverage. However, simultaneously achieving high sensitivity and broad
instantaneous bandwidth in a Rydberg-based MW sensor remains a critical
challenge. Here, we propose a multi-dressed-state engineered superheterodyne detection scheme for Rydberg electrometry to overcome this challenge. It is found that the key  to simultaneously achieving large instantaneous bandwidth and high sensitivity lies in the coherence of dressed states and the interference between transition channels of dressed states. By strategically
engineering the multiple dressed states of Rydberg atoms, we demonstrate a
thermal $\mathrm{^{87}Rb}$ vapor-based sensor with a sensitivity
of 222.6$\,$nV$\,$cm$^{-1}\,$Hz$^{-1/2}$ and a record instantaneous bandwidth
of 76.8$\,$MHz with the local microwave frequency 16.03$\,$GHz.  This advancement paves the way for Rydberg-atom technologies in radar, wireless communication, and spectrum monitoring.
\end{abstract}

\maketitle
Microwave (MW) detection plays a momentous role in the fields of
cosmological research, radar technology and satellite communications \cite{liddle2015introduction,durrer2015cosmic,skolnik2002role,pan2020microwave,sobol1984microwave,tsui2002digital,fan2015atom, yuan2023quantum,schlossberger2024rydberg}. Rydberg atoms, characterized by their high major quantum numbers and large transition dipole moments, endow Rydberg MW sensors
with significant advantages over traditional antennas, including a broad
frequency response range, compact structural dimensions, and exceptional
sensitivity \cite{saffman2010quantum,holloway2014broadband,holloway2017atom,holloway2017electric,anderson2020rydberg,meyer2020assessment,meyer2021waveguide}. These attributes have propelled Rydberg microwave detection
technology to the forefront of high-sensitivity microwave measurement
research. Sensitivity and instantaneous bandwidth (IB) are pivotal parameters in microwave measurement. Over the past decade, Rydberg MW
transducers have achieved significant advancements in optimizing these two
parameters independently \cite{sedlacek2012microwave,sedlacek2013atom,anderson2014two,miller2016radio,anderson2016optical,jing2020atomic,prajapati2021enhancement,liu2022continuous,zhang2022rydberg,liu2022deep,ding2022enhanced,tu2022high,ouyang2023continuous,berweger2023closed,borowka2024continuous,tu2024approaching,hu2023improvement,dixon2023rydberg,artusio2024increased,yang2024highly,wu2024nonlinearity,cloutman2024polarization}. A Rydberg MW transducer with a sensitivity of 55$\,$nV$\,$cm$^{-1}\,$Hz$^{-1/2}$ and a 150$\,$kHz IB using the superheterodyne method has been realized \cite{jing2020atomic}. The sensitivity has been improved to 10$\,$nV$\,$cm$^{-1}\,$Hz$^{-1/2}$ with a wider IB of 4.6$\,$MHz in a cold atom system \cite{tu2024approaching}. The dual-beam excitation in a thermal cesium vapor is utilized to enhance the IB and achieved a sensitivity of 468$\,$nV$\,$cm$^{-1}\,$Hz$^{-1/2}$ with 6.8$\,$MHz \cite{hu2023improvement}. The NIST team demonstrated a 12$\,$MHz IB with 1.91$\,\mu$V$\,$cm$^{-1}\,$Hz$^{-1/2}$ sensitivity via frequency-comb probing \cite{dixon2023rydberg,artusio2024increased}. Notably, Bowen Yang et al. established the current record of $\pm $10.2$\,$MHz IB while maintaining 62$\,$nV$\,$cm$^{-1}\,$Hz$^{-1/2}$ sensitivity through dual-six-wave-mixing superheterodyne detection in a $^{87}$Rb vapor \cite{yang2024highly}.
For practical MW sensing applications (e.g., radar, 5G communications), the IB should be as large as possible while satisfying the sensitivity requirement.
Simultaneously achieving high-sensitivity and broad-IB performance continues to pose fundamental technical challenges, necessitating novel methodological breakthroughs. 

In this article, we propose a multi-dressed-state engineered
superheterodyne (MDSES) scheme for Rydberg electrometry. This scheme can increase the instantaneous bandwidth while maintaining high sensitivity. Its physical essence lies in the coherence of dressed states and the interference effect  between dressed-state transition channels, we experimentally enhanced the instantaneous bandwidth up to 76.8$\,$MHz with a sensitivity of 222.6$\,$nV$\,$cm$^{-1}\,$Hz$^{-1/2}$.
\begin{figure}[t]
	\centering
	\includegraphics[width=0.95\linewidth]{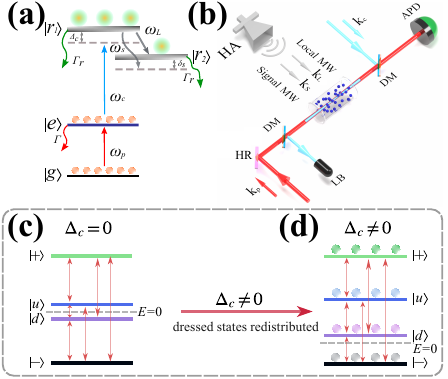}
	\caption{
		 \textbf{(a)} The atomic energy level scheme.  $\left\vert g\right\rangle$: $\left\vert {5S_{1/2}, F=2}\right\rangle $; $\left\vert e\right\rangle$: $\left\vert{5P_{3/2}, F'=3}\right\rangle $; $\left\vert r_{1}\right\rangle$: $\left\vert{51D_{5/2}, m_J=1/2}\right\rangle $; $\left\vert r_{2}\right\rangle$: $\left\vert{52P_{3/2}, m_J=1/2}\right\rangle $. $\Gamma$ is the spontaneous decay rate of $|e\rangle $, $\Gamma _{r}$ is the spontaneous decay rate of the Rydberg states $|r_{1}\rangle $ and $|r_{2}\rangle $. $\omega_{p, c, L, s}$ are the frequencies of the probe, coupling, local MW, and signal MW fields, respectively. \textbf{(b)} The experimental setup. DM: dichroic mirror; HR: high-reflection mirror; LB: laser block; HA: horn antenna; APD: avalanche photodetector. $k_{p,c,L,s}$ are the wave vectors of the probe, coupling, local MW, and signal MW fields, respectively. \textbf{(c)} The distribution of the dressed states when $\Delta_{p,c,L}=0$. \textbf{(d)} The redistribution of the dressed states when $\Delta_{p,L}=0$ and $\Delta_{c}\ne 0$. The gray dashed lines in \textbf{(c, d)} represent the zero energy reference point of the dressed states ($E=0$).
	}
	\label{fig1}
\end{figure}

The energy levels and the experimental setup are shown in Fig. \hyperref[fig1]{1(a, b)}. The atoms in a thermal $^{87}$Rb vapor cell at a temperature of 21$\,$℃ are excited from the ground state $|g\rangle$ to the Rydberg state $|r_1\rangle$ via a two-photon excitation scheme. The modulation transfer spectroscopy is used to stabilize the frequency of the 780$\,$nm probe laser, which continuously drives the atomic transition $|g \rangle \rightarrow |e \rangle$, and the 480$\,$nm coupling laser is locked to a high-finesse Fabry-P\'{e}rot cavity exciting the atoms from $|e \rangle$ to $|r_1\rangle$. The probe and coupling lights are propagating in opposite direction to reduce the Doppler effect. The waist radius of the probe and coupling beams are 52.5$\,\mathrm{\mu m}$ and 29 $\mathrm{\mu m}$, respectively. The local MW field (with a frequency of 16.03$\,$GHz and a Rabi frequency $\Omega_L$) is coupling the Rydberg states $|r_{1}\rangle $ and $|r_{2}\rangle$. The signal MW field as a perturbation is periodically driving the system with a Rabi frequency $\Omega_s$ ($\Omega_s \ll \Omega_L$) and a frequency difference $\delta_s=\omega_s-\omega_L$. The two MW fields are combined using a power combiner and emitted from a horn antenna to free space. All the fields are vertical polarized. To ensure the uniformity  of  MW, the horn antenna is positioned at a distance exceeding 50$\,$cm from the atom cell.  Furthermore,  the 5$\,$cm cell is shielded by the microwave anechoic absorbers to reduce the reflections from conductors which might otherwise influence the measurement results. The superheterodyne signal is detected by an avalanche photodetector (Thorlabs APD130A) with a bandwidth of 50$\,$MHz and finally acquired by a spectrum analyzer (Keysight N9020A). The signal generators and the spectrum analyzer are synchronized to a 10$\,$MHz reference signal derived from a high-stability crystal oscillator.

We first present a theoretical analysis of the physical mechanism underlying the IB enhancement in MDSES. As illustrated in Fig. \hyperref[fig1]{1a}, three strong fields--probe ($\omega_{p}$), coupling ($\omega_{c}$), and local MW ($\omega_{L}$) fields,  and one weak signal field ($\omega_{s}$) are applied to a four-level atomic system. The three strong fields generate four dressed states (Fig. \hyperref[fig1]{1c}), while the weak field—despite being too weak to participate in the dressed-state formation—plays a decisive role in mediating the coupling between dressed states. Specifically, the distribution of dressed states are determined by the strong field parameters, such as the Rabi frequecy $\Omega_{p,c,L}$ and the frequency detuning $\Delta_{p,c,L}$, and the signal field parameters $\Omega_s$ and $\delta_s$ affect the coupling between dressed states. We further consider the function of the signal field on the dressed states. The signal field part of the Hamiltonian in the dressed-state basis is given by
\begin{equation}
	\hat{H}_{s}=\frac{\Omega_s}{2} e^{i\delta_s t}\sum_{ij} \bm{C}_{ij}\hat{\sigma}_{ij}+ H.c.
	\label{eq1}
\end{equation}
$\hat{\sigma}_{ij}$ is the atomic transition operator between dressed states $|i\rangle$ and $|j\rangle$ ($i, j \in \{-,d,u,+\}$), and  $\bm{C}_{ij}$ (determined by $\Omega_{p,c,L}$ and $\Delta_{p,c,L}$) is the projection coefficient \cite{bracht2023dressed,li2024resonant, liu2006atom}.
According to Eq.(\ref{eq1}), the effect of the signal field on the dressed states is driving transitions between dressed states with Rabi frequency $\Omega_s$ and a frequency detuning $\delta_s$.
The measured transmission signal can be characterized by the atomic coherence $\rho_{ge}$ \cite{Cohen-Tannoudji1996,Boyd2020Nonlinear, PhysRevA.88.033417}. Projecting $\rho_{ge}$ to the dressed-state basis, it involves all the dressed-state transition channels.  Consequently, the response amplitude is governed by the six dressed-state transition channels, which is given by (details in the Supplementary Materials)
\begin{equation}
	\left\vert\tilde{S}(\delta_s)\right\vert=\sqrt{\sum_{k}|\tilde{S}_k(\delta_s)|^2+\sum_{\alpha \beta}|\tilde{S}_\alpha(\delta_s)||\tilde{S}_\beta(\delta_s)|\cos\Delta\phi_{\alpha \beta}}
	\label{eq2}
\end{equation}
\begin{figure}[b!]
	\centering
	\includegraphics[width=0.45\textwidth]{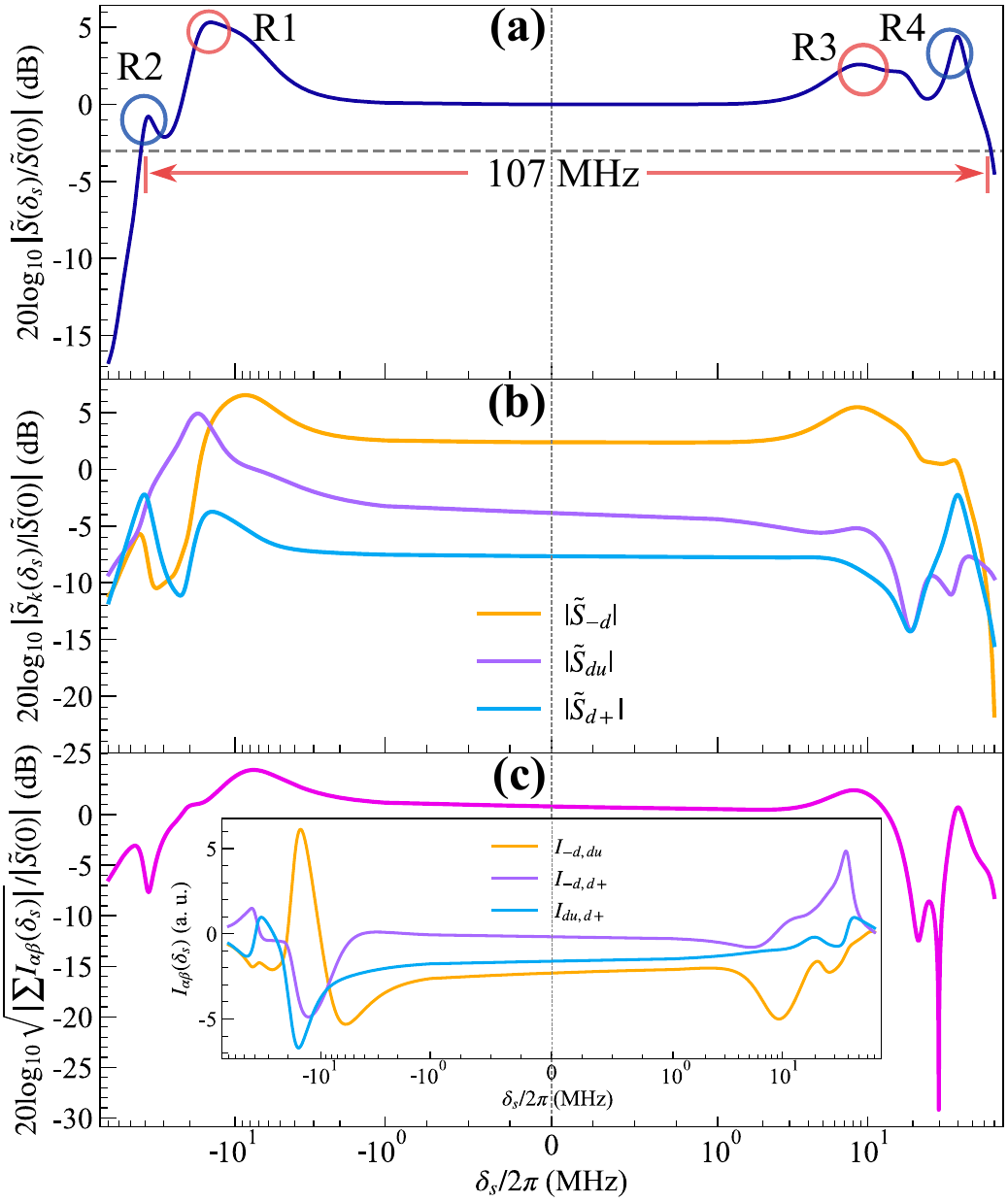}
	\caption{The theoretical results calculated by Eq. (\ref{eq2}) at $\Delta_c/2\pi=-30\,$MHz when $\Omega_p/2\pi=28.5\,$MHz, $\Omega_c/2\pi=171\,$MHz, and $\gamma t/2\pi=3.5\,$MHz, where $\gamma t$ is the transit relaxation rate. \textbf{(a)} The normalized overall response amplitude. The gray dashed line represents the 3$\,$dB attenuation of the response amplitude. \textbf{(b)} The three dominant dressed-state atomic coherence as a function of $\delta_s$.  \textbf{(c)} The contribution of interference effect to the overall response amplitude. For a clear observation of line-shape evolution, all the theoretical results are normalized to $|\tilde{S}(0)|$. The three dominant dressed-state transition interference effects are shown in the inset.
	}
	\label{fig2}
\end{figure}
\begin{figure*}[t!]
	\centering
	\includegraphics[width=0.8\textwidth]{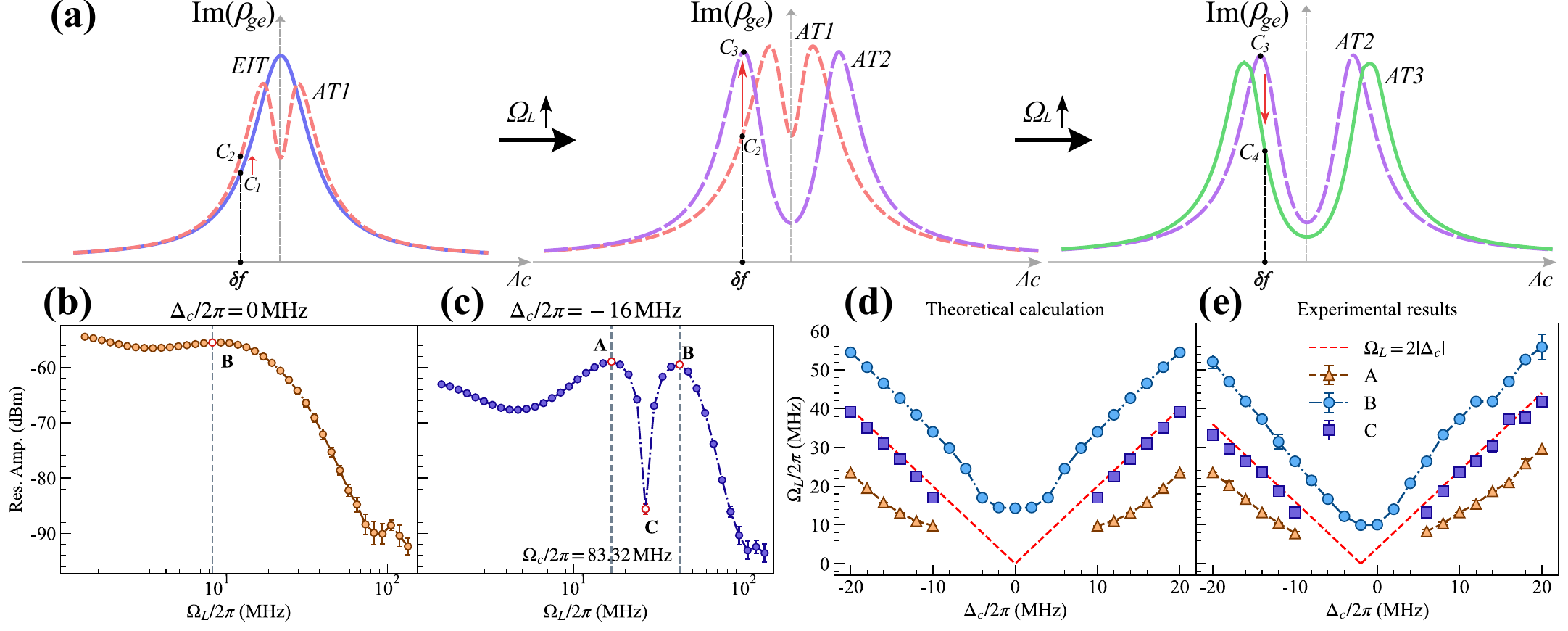}
	\caption{
		The physical mechanism for dual response
		peaks, the theoretical calculations, and the experimental results. \textbf{(a)} Physical mechanisms for
		dual response peaks A and B. $\Delta _{c}$ is fixed at $\delta f$ detuning and $\Omega _{L}$ increase sequentially from left to right. $\rho_{ge}$: the atomic coherence between $\left\vert g\right\rangle$ and $\left\vert e\right\rangle$. AT: Autler-Townes splitting. EIT: electromagnetically induced transparency. \textbf{(b, c)} The response amplitude as a function of $\Omega_{L}$ at $\Delta_{c}/2\pi =0\,$MHz and $\Delta_{c}/2\pi =-16\,$MHz. (\textbf{d, e)} $\Omega _{L}$ as a function of $\Delta _{c}$ at response peaks A and B in theory  and experiment. The red dashed lines represent $\Omega _{L}=2|\Delta _{c}|$, where the slope of Autler-Townes splitting is zero ($C_3$) as shown in \textbf{(a)}. The signal MW power was kept at -45$\,$dBm and $\left\vert \delta _{s}\right\vert /2\pi =200\,$kHz.	} 
	\label{fig3}
\end{figure*}
where $k, \alpha, \beta \in \left\{| i \rangle \leftrightarrow | j \rangle\right\}$ represent the different dressed-state transition channels, $|\tilde{S}_{k, \alpha, \beta}(\delta_s)|$ is the oscillation amplitude in the imaginary part of the coherence  between dressed states $ | i \rangle $ and $| j \rangle$, $\Delta \phi_{\alpha \beta}=\phi_\alpha-\phi_\beta$ is the phase difference between $\tilde{S}_{\alpha}(\delta_s)$ and $\tilde{S}_{\beta}(\delta_s)$, and $\phi_{\alpha, \beta}=\text{arg}\left[\tilde{S}_{\alpha, \beta}(\delta_s)\right]$ is the additional phase induced by the signal MW field. To determine IB, we maintain $\Omega_s$ as a constant, and increase $|\delta_s|$ from zero to obtain the curve of $|\tilde{S}(\delta_s)|$. We define the normalized response amplitude as $\tilde{S}_N(\delta_s)=20\log_{10}\left\vert\tilde{S}(\delta_s)/\tilde{S}(0) \right\vert$, the IB is taken as  the sum of the positive and negative $\delta_s$ where $\tilde{S}_N(\delta_s)$ falls to $-3\,$dB. Eq. (\ref{eq2}) reveals that the response amplitude has two physical origins, one is the dressed-state coherence ($|\tilde{S}_{k}|$), another is the interference between dressed-state transition channels ($I_{\alpha\beta}(\delta_s)=|\tilde{S}_{\alpha}(\delta_s)| |\tilde{S}_{\beta}(\delta_s)|\cos\Delta\phi_{\alpha \beta}$).

\begin{figure}[h!]
	\centering
	\includegraphics[width=0.45\textwidth]{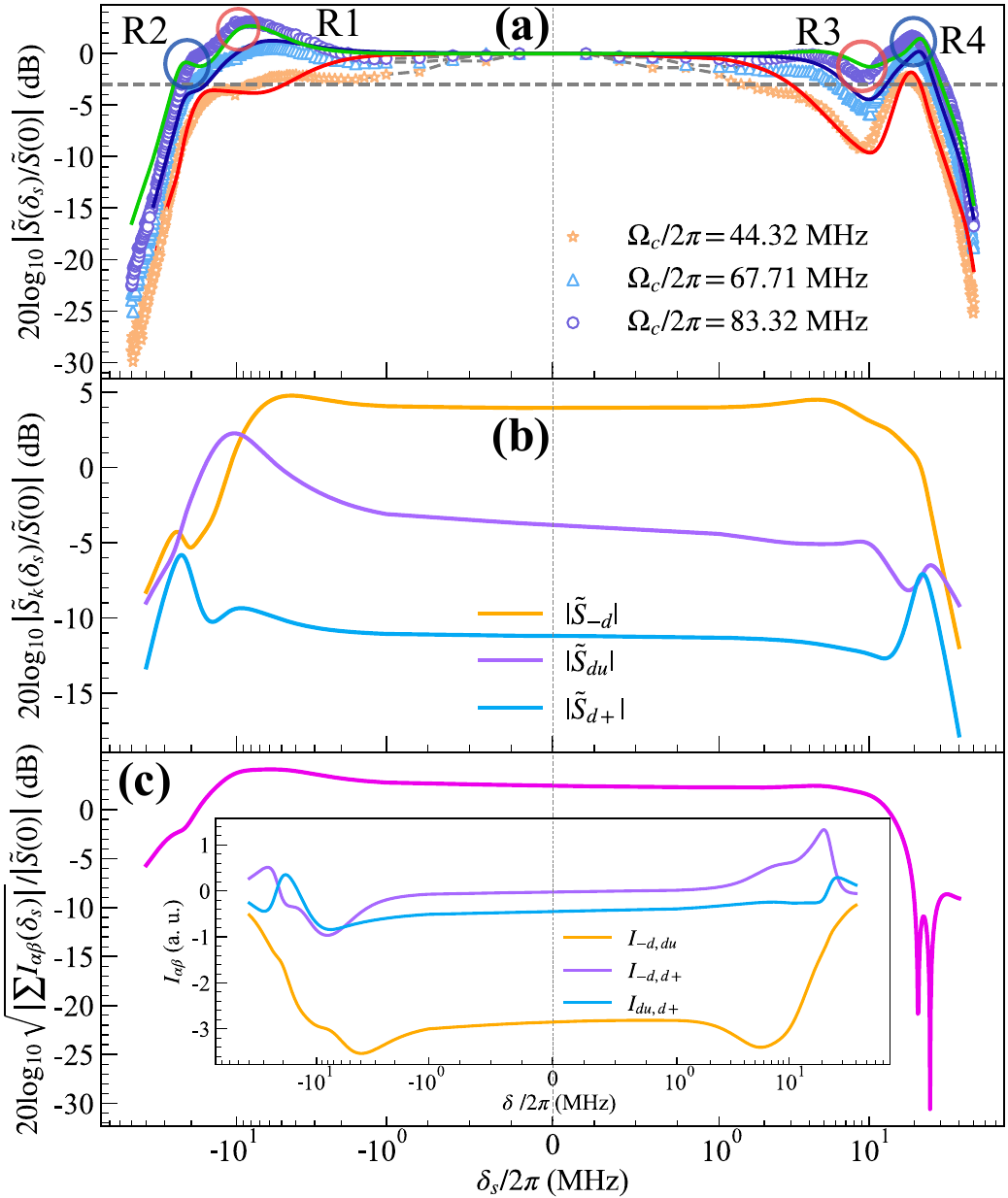} 
	\caption{
		The experimental frequency response curves at $\Delta_c/2\pi=-16\,$MHz and the theoretical analysis.
		\textbf{(a)} The frequency response curves at three different $\Omega_c$. The solid lines represent the theoretical calculations derived from Eq. (\ref{eq2}), including the contribution from the dressed-state atomic coherence and the interference between different transition channels. The gray dashed line represents the 3$\,$dB attenuation of the response amplitude relative to the value $|\delta_s|/2\pi=200\,$kHz. The response amplitude is normalized to that of $|\delta_s|/2\pi=200\,$kHz. \textbf{(b)} The theoretical results of three dominant dressed-state coherences with $\Omega_c/2\pi=83.32\,$MHz. \textbf{(c)} The theoretical contribution of interference effect to the overall response amplitude with $\Omega_c/2\pi=83.32\,$MHz. For a clear observation of line-shape evolution, all the theoretical results are normalized to $|\tilde{S}(0)|$.  The three dominant dressed-state transition interference effects are shown in the inset. 
	} 
	\label{fig4}
\end{figure}

Both the coherence between dressed-states and interference between different dressed-state transition channels can be manipulated by tuning the frequency detuning $\delta_s$. When $|\delta_s|$ is resonant with two dressed states, the coherence between them reaches maximum, and a relatively strong interference effect between different transition channels needs $|\delta_s|$ at resonance or near-resonance with these channels, thereby endowing the atomic transitions between the dressed states with strong competition. 
In most previous experimental and theoretical studies, the detuning was typically set to zero ($\Delta_{p,c,L}=0$), resulting in a symmetric distribution of the dressed states, as shown in Fig. \hyperref[fig1]{1c}. The frequency differences among the six coherent transition channels are relatively large. Consequently, when the incident signal field detuning $|\delta_s|$ is resonant with a specific transition channel (e.g., $|d \rangle \leftrightarrow |u\rangle$), the system exhibits high sensitivity. However, as $|\delta_s|$ gradually deviates from the resonance frequency, the coherences among the other dressed states become difficult to establish effectively. As a result, the response amplitude $|\tilde{S}(\delta_s)|$ decreases, leading to a degradation of IB. For avoiding this degradation, we introduce a coupling light detuning $\Delta_c \ne 0$, which breaks the symmetric distribution of the dressed states, causing their energy levels to shift asymmetrically with respect to $E = 0$, as shown in Fig. \hyperref[fig1]{1d}.
The direction of the shift is determined by whether the coupling field is red- or blue- detuned. 
This redistribution of the dressed-state energy levels makes the frequency differences among the six coherent transition channels near to each other. Therefore, when the signal microwave frequency gradually detunes away from a resonant transition channel, the contributions from the other coherent transition channels increase rapidly, enabling mutual compensation of coherences and thereby effectively suppressing the rapid degradation of the IB performance.
Therefore, the introduction of $\Delta_c$ modifies the dressed-state distribution and the projection coefficient $\bm{C_{ij}}$, and alters the coherences between dressed states and the interference between transition channels during the tuning of $\delta_s$. As shown in Fig. \hyperref[fig1]{1d} for the dressed-state energy levels, when we increase $|\delta_s|$ from zero, the coherence between $| - \rangle$ and $| d \rangle$ will be established first, and $|\tilde{S}_{-d}|$ dominant the response. When $|\delta_s|$  is detuned away from $\omega_{-d}$, $|\tilde{S}_{-d}|$ decays. However, since the resonant frequency $\omega_{-d}$ and  $\omega_{du}$ are not significant different, $|\tilde{S}_{du}|$ can compensate for the decay of $|\tilde{S}_{-d}|$. At the same time, the interference between the dressed-state transition channels should be considered. This complementary coherence and interference effects prevent rapid decay of the overall response amplitude and degradation of IB. We analyzed a configuration yielding an IB of 107$\,$MHz with $\Omega_p/2\pi=28.5\,$MHz and $\Omega_c/2\pi=171\,$MHz (all the Rabi frequencies in the following theory and experiment are the peak Rabi frequencies). The overall response amplitude are presented in Fig. \hyperref[fig2]{2a}. The response exhibits asymmetric patterns at the $\delta_{s}>0$ and $\delta _{s}<0$ regimes due to the asymmetric dressed-state distribution.  We also calculated the dominant dressed-state coherences and the contribution of interference effect to the overall response amplitude, which are presented in Fig. \hyperref[fig2]{2(b,c)}, respectively, with the three dominant interference components are shown in the inset of Fig. \hyperref[fig2]{2c}.
It can be observed that the peak shapes of the overall response curve primarily originate from the coherence effects of three pairs of dressed states ($|\tilde{S}_{-d}|$, $|\tilde{S}_{du}|$, $|\tilde{S}_{d+}|$), and the interference effects among three dressed-state transition channels ($I_{-d,du}$, $I_{-d,d+}$, $I_{du,d+}$). The peak shapes $R1$ and $R2$ mainly stem from the contributions of coherence, whereas the peak shapes $R3$ and $R4$ arise from the combined contributions of both coherence and channel interference effects. As $|\delta_s|\rightarrow\infty$, the coupling between dressed states is suppressed, giving the response amplitude $|\tilde{S}(\delta_s)|\rightarrow 0$. Subsequent experimental measurements realize the state-of-the-art IB, and the theoretical results explain the profile of the experimental response curves by the MDSES theory.

In our experiment, we set $\Omega_p/2\pi=17.16\,$MHz and $\Delta_p=\Delta_L=0$.
The system initially optimizes the $\Omega_L$ to maximize the response amplitude. Fig. \hyperref[fig3]{3} shows the superheterodyne response's dependence on $\Omega_{L}$ under fixed $\Delta _{c}$ and $|\delta _{s}|/2\pi=200\,$kHz. For resonant coupling ($\Delta _{c}=0\,$MHz, Fig. \hyperref[fig3]{3b}), the response curves exhibit a single-peak characteristic with the increase of $\Omega _{L}$. When $\Delta _{c}\neq 0\,$MHz (Fig. \hyperref[fig3]{3c}), the detuning frequency induces a bifurcation into two spectrally distinct peaks, denoted as peak A and peak B, whose physical origin is elucidated in Fig. \hyperref[fig3]{3a}. When $\Delta _{c}$ is fixed at a detuning frequency $\delta f$ ($\left\vert\delta f\right \vert>\Omega _{L}/2$), progressive increase of $\Omega _{L}$ generates monotongeically increasing Autler-Townes (AT) splitting, driving the $\text{Im}\left[\rho_{ge}\left(\delta f\right)\right]$ through critical points $C_{1}$--$C_{4}$. The dual-peak structure emerges from slope extrema at $C_{2}$ (forming peak A) and $C_{4}$ (forming peak B), separated by a null-slope inflection at $C_{3}$.
It is known that 
$\Omega_L$ must be greater than $\Gamma_{EIT}/2$ for AT splitting to be observed, where $\Gamma_{EIT}$ is the full width at half maxima (FWHM) of the EIT spectrum. Moreover, according to the physical mechanism illustrated in Fig. \hyperref[fig3]{3a}, the emergence of peak A requires that $\text{Im}\left[\rho_{ge}\left(\delta f\right)\right]$ experience $C_2$ during the increase of 
$\Omega_L$. Therefore, peak A can only be observed if $|\Delta_c|/2\pi>\Gamma_{EIT}/2$. When the system is near resonance, it only experiences $C_4$ during the increase of $\Omega_L$, producing a single response maximum at $C_4$ instead of the dual-peak structure.
As illustrated in Fig. \hyperref[fig3]{3e}, the response peak A disappears when $-10\,\mathrm{MHz} <\Delta _{c}/2\pi < 6\,\mathrm{MHz}$, resulting in the emergence of a single peak B. The experimental results have a 2$\,$MHz offset compared to the theoretical calculations (Fig. \hyperref[fig3]{3d}). This offset arises from the AC Stark shift of the Rydberg energy levels induced by the strong coupling optical field and local MW field \cite{bracht2023dressed,liu2024higher}, and the Zeeman splitting induced by the geomagnetic field. Considering the energy level shift, the behavior that peak A vanishes for $-8\,\mathrm{MHz} <\Delta _{c}/2\pi < 8\,\mathrm{MHz}$ satisfy the condition of $|\Delta _{c}|/2\pi<\Gamma_{EIT}/2=8.63\,$MHz. 

To quantify the IB of the system, we measured the frequency response curves at different  $\Delta_c$ ($\Delta_p=\Delta_L=0$). For example, with $\Delta_c/2\pi=-16\,$MHz and  $\Omega_L/2\pi=16.66\,$MHz (peak A) is shown in Fig. \hyperref[fig4]{4a}.  When $\Omega _{c}/2\pi=83.32\,$MHz, the IB is 31.2$\,$MHz for $\delta_s<0$ and 20.4$\,$MHz for $\delta_s>0$. The solid lines in Fig. \hyperref[fig4]{4a} correspond to theoretical results derived from Eq. (\ref{eq2}), demonstrating good agreement with the experimental measurements.
When $\delta_s<0$, as shown in Fig. \hyperref[fig4]{4b}, the coherence between $| - \rangle$ and $| d \rangle$ dominates the curve profile in the low-$|\delta_s|$ regime.
As $|\tilde{S}_{-d}|$ is approaching its peak value ($|\delta_s|$ at resonance of $|-\rangle \leftrightarrow |d\rangle$), $|\tilde{S}_{du}|$ is also rising (since $\omega_{-d}\approx \omega_{du}$), forming the red-circled region $R1$ in Fig. \hyperref[fig4]{4a}. We also observed that the response amplitude in the blue-circled region $R2$ of Fig. \hyperref[fig4]{4a}. This is because $|\delta_s|$ is near the resonant frequency of $|d\rangle \leftrightarrow |+\rangle$, $|\tilde{S}_{d+}|$ rises gradually to compensate for the decay of $|\tilde{S}_{-d}|$ and $|\tilde{S}_{du}|$. When $\delta_s>0$, $|\tilde{S}_{-d}|$ consistently makes the dominant contribution. The dip $R3$ is primarily attributed to the decrease of $|\tilde{S}_{-d}|$. As $|\delta_s|$ is approaching the resonant frequency of $|d\rangle \leftrightarrow |+\rangle$, $|\tilde{S}_{d+}|$ reaches its peak, and there is a constructive interference between the transition channels $|-\rangle \leftrightarrow |d\rangle$ and $|d\rangle \leftrightarrow |+\rangle$. These combined effects result in the emergence of the peak-shaped feature $R4$ in the overall response curve. We demonstrated the feasibility of MDSES both experimentally and theoretically.

\begin{figure}[t!]
	\centering
	\includegraphics[width=0.48\textwidth]{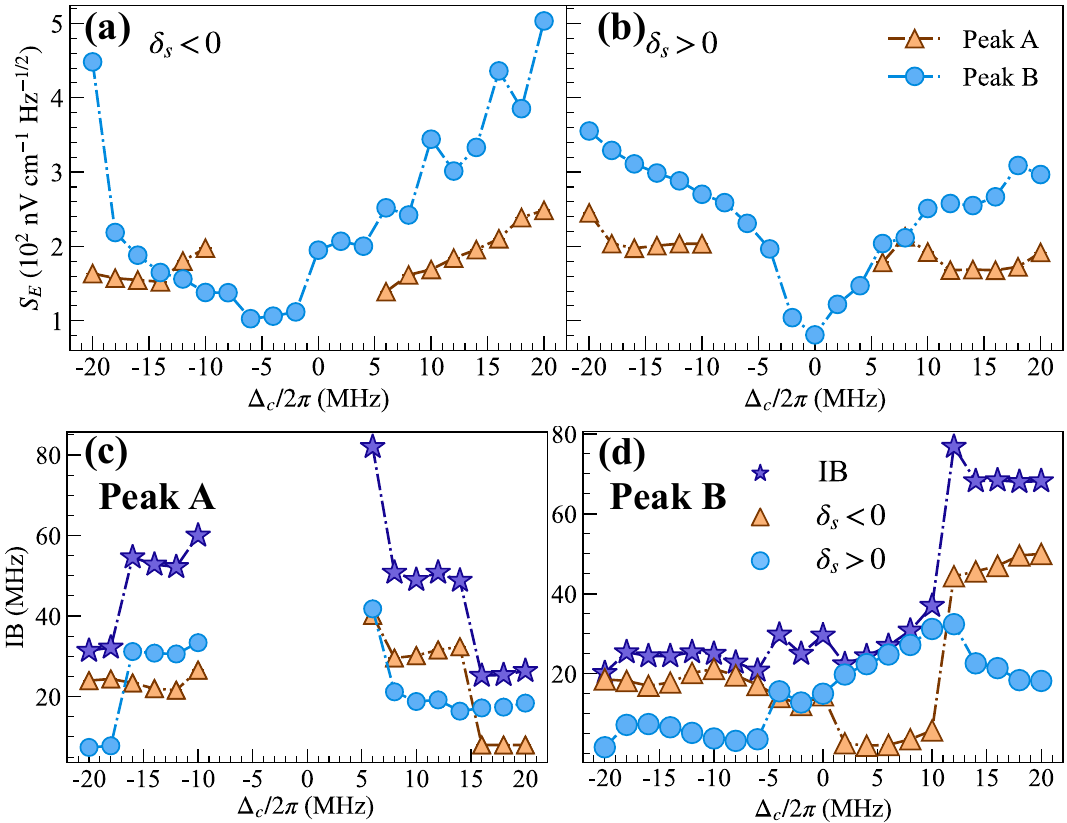} 
	\caption{
		Experimental results of IB and sensitivity. \textbf{(a, b)} The optimal sensitivity at different $\Delta_c$ when $\delta_s<0$ \textbf{(a)} and $\delta_s>0$ \textbf{(b)}. \textbf{(c, d)}  The IB when $\Omega_L$ at peak A \textbf{(c)} and $\Omega_L$ at peak B \textbf{(d)} at different $\Delta_c$. The IB was measured while maintaining the signal MW power at --45$\,$dBm. All the experimental results were measured when $\Omega_p/2\pi=17.16\,$MHz, $\Omega_c/2\pi=83.32\,$MHz, and $\Delta_p=\Delta_L=0\,$MHz.
	}
	\label{fig5}
\end{figure}

The experimental data in Fig. \hyperref[fig5]{5(a-d)} present the sensitivity and IB of the Rydberg MW sensor at different $\Delta_c$. For $\Omega_L/2\pi=16.66\,$MHz (peak A) and $\Delta_c/2\pi = -16\,$MHz, the IB reaches 54.6$\,$MHz with an optimal sensitivity of 140.4$\,$nV$\,$cm$^{-1}\,$Hz$^{-1/2}$. The maximum IB of 82$\,$MHz occurs at $\Delta_c/2\pi = 6\,$MHz with $\Omega_L/2\pi=8.36\,$MHz (peak A), however exhibits drastically reduced sensitivity to 1.25$\,\mathrm{\mu}$V$\,$cm$^{-1}\,$Hz$^{-1/2}$ at $|\delta_s|/2\pi=200\,$kHz (see Supplementary Materials). The best sensitivity achieves 81.1$\,$nV$\,$cm$^{-1}\,$Hz$^{-1/2}$ at $\Delta _{c}$ = 0$\,$MHz with $\Omega_L/2\pi=10.05\,$MHz (peak B)  as shown in Fig. \hyperref[fig5]{5d}. At $\Delta_c/2\pi = 12\,$MHz and $\Omega_L/2\pi=41.84\,$MHz (peak B), the optimal performance of the Rydberg MW sensor is realized, where the IB reaches 76.8$\,$MHz with an optimal sensitivity of 222.6$\,$nV$\,$cm$^{-1}\,$Hz$^{-1/2}$. 

In summary, we demonstrated a multi-dressed-state engineered superheterodyne (MDSES) method to simultaneously achieve high sensitivity and broad instantaneous bandwidth for Rydberg electrometry. The underlying physical mechanism of this scheme stems from the coherence of dressed states and the quantum interference between the transition channels of these dressed states. Using this method, we achieved an instantaneous bandwidth of 76.8$\,$MHz and a sensitivity of 222.6$\,$nV$^{-1}\,$cm$^{-1}$Hz$^{-1/2}$.
The instantaneous bandwidth can be boosted to over 100$\,$MHz in theory, and the microwave signal frequency can be extended to  other bands. Our results establish a universal pathway toward the practical deployment of high-performance Rydberg microwave sensors. The demonstrated dressing-based quantum control paradigm also holds great promise for broader applications in quantum-enhanced metrology, especially for other atomic systems where analogous engineering of quantum state coherence is feasible.
\newline

This work is supported by the National
Key Research and Development Program of China
(2024YFA1409404), the National Natural Science Foundation of China (Grants No. 12174409, No. 12304294, No. U23A2075, and No. 12274132), the Fundamental
Research Funds for the Central Universities and Shanghai
Municipal Education Commission (202101070008E00099).
\newline

%\nocite{*}
\twocolumngrid
\bibliography{BandwidthArticle}% Produces the bibliography via BibTeX.

\end{document}